\documentclass[amsmath,amssymb,twocolumn,prl,10pt,showpacs]{revtex4}
\usepackage{graphics}
\usepackage{dcolumn}
\usepackage{bm}
\begin{document}
\author{Bj\"orn Hessmo$^{1}$}
\email{hessmo@imit.kth.se}
\author{Pavel Usachev$^{2}$}
\affiliation{}
\author{Hoshang Heydari$^{1}$}
\author{Gunnar Bj\"ork$^{1}$}
\affiliation{%
$^{1}$Department of Microelectronics and Information Technology, Royal
Institute of Technology (KTH), S-16440 Kista, Sweden\\
$^{2}$Ioffe Physico-technical Institute, Russian Academy of
Sciences, Politekhnicheskaya ul. 26, St. Petersburg, 194021
Russia}
\newcommand{\infig}[2]{\begin{center}\resizebox{#2}{!}{\includegraphics*{#1}}\end{center}}
\newcommand{\figtext}[1]{{#1}}
\newcommand{\ket}[1]{\left| #1 \right\rangle}
\newcommand{\bra}[1]{\left\langle #1 \right|}
\newcommand{\Adagger}{\hat{a}^{\dag}}
\newcommand{\Bdagger}{\hat{b}^{\dag}}
\newcommand{\A}{\hat{a}}
\newcommand{\B}{\hat{b}}

\title{An experimental demonstration of single photon nonlocality}
\begin{abstract}
In this letter we experimentally implement a single photon Bell
test based on the ideas of S. Tan \emph{et al.} [Phys. Rev. Lett.,
\textbf{66}, 252 (1991)] and L. Hardy [Phys. Rev. Lett.,
\textbf{73}, 2279 (1994)]. A double heterodyne measurement is used
to measure correlations in the Fock space spanned by zero and one
photons. Local oscillators used in the correlation measurement are
distributed to two observers by co-propagating it in an orthogonal
polarization mode. This method eliminates the need for
interferometrical stability in the setup, consequently making it a
robust and scalable method.
\end{abstract}

\maketitle

For experimental Bell tests \cite{Bell1964,Clauser1969} it has
been a successful strategy to use polarization entangled photon
pairs, either from atomic cascades
\cite{Freedmann1972,Aspect1981,Aspect1982a,Aspect1982b}, or
parametric down-conversion
\cite{Ou1988,Kwiat1995,Weihs1998,Kurtsiefer2001}, or produced by
post selecting a photon pair from independent sources
\cite{Pittman2003}. In these experiments, it is observed that
correlations between the two photons are incompatible with local
realism, i.e. Bell's inequalities are violated.

In 1991, Tan \textit{et al.} proposed that it would be possible to
show a contradiction between local realism and quantum mechanics using
only a single particle \cite{Tan1991}. The proposal spurred a
debate, where the main argument against the feasibility of such an
experiment was that detection of the particle at one location
would prohibit the measurement of any property associated with
that particle at another location. A counter argument was
that this would indeed be true if one measured particle-like
properties at the two locations, but if one instead chose to
measure wave-like properties, the argument fails. Another
criticism against Tan's \textit{et al.} proposal was that a
measurement of wavelike properties requires a reference
oscillator. Hence, measurements of wavelike properties of a single
particle require additional particles.  The notion of
\textit{single} particle non-locality was hence put in doubt
\cite{Greenberger1995comment}. Hardy, who had proposed an
alternative experiment to demonstrate the non-locality of a single
particle, then put forth operational criteria for such a test,
where the main ingredient is that the demonstrated non-local
properties should be dependent on the presence of a single
particle \cite{Hardy1994,Hardy1995reply}. If the quantum state is
robbed of this particle, no non-local properties should be
observed. That is, all the non-local correlations should be
carried by a single-particle state, although the observation of
these correlations may contain measurements on auxiliary reference
particles. The state of these reference oscillators should be such
that the observers can generate them using only classical
communication and local operations.

In this letter we experimentally test the behavior of non-local
correlations for one single photon using a setup similar to the
one proposed by Tan \emph{et al.} \cite{Tan1991}. When a single
photon passes a balanced beam splitter it has equal probability
amplitudes for reflection and transmission. In the number basis
such states have the form
\begin{equation}
\frac{1}{\sqrt{2}}\left[ |1\rangle_\mathrm{T}|0\rangle_\mathrm{R}+
|0\rangle_\mathrm{T}|1\rangle_\mathrm{R} \right],
\label{eq:bellstate}
\end{equation}
where T (R) refers to the transmitted (reflected) channel of the
beam splitter. This state is mathematically isomorphic to a
two-photon Bell state encoded in horizontal (H) and vertical (V)
polarization, with the replacements
$|0\rangle\leftrightarrow|H\rangle$ and
$|1\rangle\leftrightarrow|V\rangle$). Using (\ref{eq:bellstate})
for a Bell experiment requires that measurements can be made in
bases complementary to the number basis $|0\rangle$ and
$|1\rangle$ in the two channels. This is not straightforward using
photon counters, since quantities complementary to photon number
are sought.

\begin{figure}[htbp]
\infig{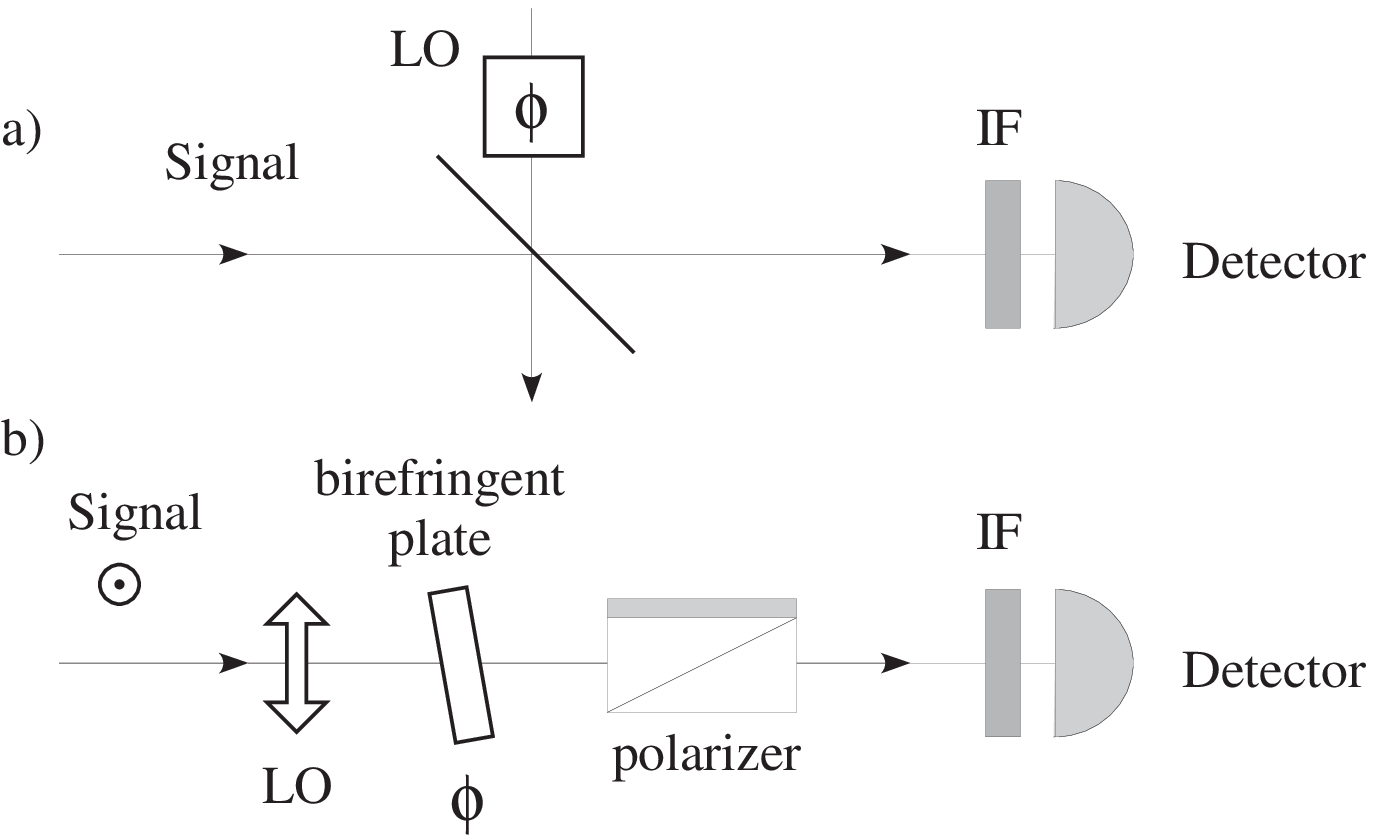}{8cm} \caption{The photon counters are
unable to tell if a detected photon came from the local oscillator
(LO) or if it was a part of the signal from the experiment. If the
two possible origins cannot be told apart, the amplitudes for the
two possible events must be added. In a) the local oscillator is
mixed with the signal on a beam splitter with high transmission.
The polarization of the signal and LO are the same to ensure
undistinguishable photons. In b) the polarization of the local
oscillator is orthogonal to the signal beam, and the two beams
co-propagates. By rotating a polarizer one can adjust the
probability for transmitting the signal, or LO-photon. In both
cases, interference filters are put before the detectors to erase
path information stored in the energy spectrum of the photons.}
\label{fig:detectors}
\end{figure}

In Fig.~\ref{fig:detectors} we illustrate two experimental
implementations capable of performing measurements complementary
to photon number measurement.

A signal from an experiment is mixed with a local oscillator (LO)
on a beam splitter in such a way that a photon detector observing
one photon is not capable of telling if the photon came from the
experiment or the LO beam. If the photon came from the LO, there
was zero photons coming from the experiment. If no photon came
from the LO, there was one photon coming from the experiment. If
the local oscillator is a coherent state $\ket{\alpha}$,
amplitudes for these two events give projection on the wanted
state:
\begin{equation}
\mathcal{N}\left(r\alpha\ket{0}+\ket{1}\right),
\label{eq:projstate}
\end{equation}
where $\mathcal{N}$ is a normalization constant, and $r\alpha$ is
the complex amplitude for the LO in the counter mode. This is
illustrated in Fig.~\ref{fig:detectors}(a) where $r\alpha$ is the
amplitude for reflection of the local oscillator and in
Fig.~\ref{fig:detectors}(b) where $r\alpha$ is the amplitude of
the LO after the polarizer. The number states in
Eq.~(\ref{eq:projstate}) describe the photon number in the mode
arriving from the experiment. For similar implementations where
local oscillators are replaced by single photons we refer to the
papers by F. Sciarrino \emph{et al.} \cite{Sciarrino2002} and H.
Lee \emph{et al.} \cite{Lee2000}.

\begin{figure}[htb]
 \infig{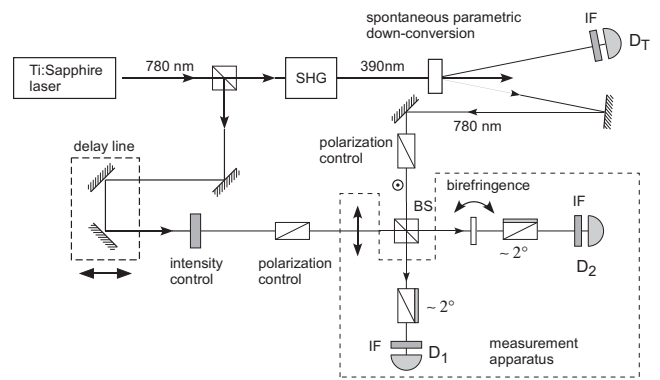}{8.5cm}
   \caption{Experimental setup: A single photon produced in
   spontaneous parametric down-converter is overlapped with a
   coherent local oscillator on a 50/50 beam splitter.
   In each output mode of the beam splitter we place the
   detectors described in Fig.~\ref{fig:detectors}.
   To change the relative phase of the local oscillators we
   tilt a birefringent quartz plate in one of the arms.}
   \label{fig:setup}
\end{figure}

In Fig.~\ref{fig:setup} we have a schematic description of the
optical setup used for this experiment. The light source is a
Ti:Sapphire laser which pumps a frequency doubler (1 mm LBO
crystal), marked SHG in the figure, producing fs-pulses at 390 nm.
These pulses pump a type-I down-converter (3 mm BBO crystal),
yielding photon pairs at 780 nm emitted with a separation of 3
degrees from the pump beam. One of the photons (the idler) is sent
through an interference filter and detected by an avalanche
photodiode, $\mathrm{D}_\mathrm{T}$. Detection of the idler in
$\mathrm{D}_\mathrm{T}$ indicates that the other photon of the
pair (the signal) is present in the experiment. The polarization
of the signal photon is made strictly vertical by a wave-plate and
a polarizer. Afterwards, the signal photon propagates to one of
the beam splitter input ports (BS).

To generate the local oscillator (LO) some light is picked off the
main beam of the Ti:Sapphire laser. A delay line is adjusted so
that the LO arrives at the beam splitter simultaneously with the
single photon. Before the beam splitter the intensity is adjusted
so that $r\alpha$ matches the single photon intensity to ensure
high visibility in the correlation measurement (See discussion
following Eq.~\ref{eq:coincfalse}).

The polarization of the LO is adjusted to be strictly orthogonal
to the signal photon polarization with another wave-plate and
polarizer. After the beam splitter each arm is equipped with a
polarizer oriented so that it transmits a LO photon or a single
photon with equal probability. Because the LO has much higher
intensity than the single photon beam the polarizer is set around
2 degrees. The single photon and the local oscillator
co-propagates after the beam splitter. This eliminates the need to
stabilize the relative phase of the local oscillators in the two
arms \cite{Bjork2002}. This is the main difference between our
implementation and the setup proposed in \cite{Tan1991}. The
relative phase between the two local oscillators is adjusted by
tilting a thin quartz plate around its optics axis in one arm. The
optic axis of the quartz plate is parallel to the single photon
polarization.

After the polarizers the light is coupled into single mode optical
fibers preceded by interference filters (FWHM 3nm) and followed by
silicon avalanche photodiodes. The signals from the APDs are
correlated and recorded by a computer.

In the optical setup illustrated in Fig.~\ref{fig:setup} the
single photon enters the beam splitter (BS) from above and the LO
photon from the left. This state is
$\ket{1,0}_{\mathrm{signal}}\otimes\ket{0,\sqrt{2}\alpha}_{\mathrm{LO}}$,
where the two modes at each location refer to the different
polarizations of the single photon channel and the local
oscillator, respectively. After the beam splitter the state is
(ignoring phase factors obtained upon reflection):
\begin{equation}
\label{eq:state}
\ket{\Psi}=\frac{1}{\sqrt{2}}(\ket{1,\alpha}_\mathrm{T}\ket{0,\alpha'}_\mathrm{R}
                            +\ket{0,\alpha}_\mathrm{T}\ket{1,\alpha'}_\mathrm{R}),
\end{equation}
where the subscripts $T$ and $R$ refer to the transmitted and
reflected output arms, respectively. The coherent states are
described by $\alpha=|\alpha|e^{i\theta}$ and
$\alpha'=|\alpha|e^{i\theta'}$. The two modes in each arm refer to
polarization. We define creation operators for
these modes as:

\begin{eqnarray*}
\Adagger_\mathrm{k}\ket{0,0}_\mathrm{k}&=&\ket{1,0}_\mathrm{k},\\
\Bdagger_\mathrm{k}\ket{0,0}_\mathrm{k}&=&\ket{0,1}_\mathrm{k},
\end{eqnarray*}
where the subscript $\mathrm{k}$ refers to the two channels R and
T. The state (\ref{eq:state}) is analyzed using a polarizer
described by the transmittance $t$ and the reflectance $r$. The
unitary transformation for this device relates the transmitted
(reflected) mode $c$ ($d$) to the incoming modes $a$ and $b$ in
the following way:

\begin{eqnarray*}
\begin{array}{cc}
\left\{
  \begin{array}{l}
    c_\mathrm{k}^\dag=ta_\mathrm{k}^\dag+rb_\mathrm{k}^\dag\\
    d_\mathrm{k}^\dag=-ra_\mathrm{k}^\dag+tb_\mathrm{k}^\dag
  \end{array}
\right. &\Leftrightarrow \left\{
  \begin{array}{l}
    a_\mathrm{k}^\dag=tc_\mathrm{k}^\dag-rd_\mathrm{k}^\dag \\
    b_\mathrm{k}^\dag=rc_\mathrm{k}^\dag+td_\mathrm{k}^\dag
  \end{array}
\right.
\end{array}
\end{eqnarray*}
Later, we will be interested in the transmitted channels
$c_\mathrm{k}$, where our detectors are placed. Rewriting the
state (\ref{eq:state}) using the above creation operators we have:
\begin{eqnarray}
\label{eq:thestate}
\ket{\Psi}&=&\frac{1}{\sqrt{2}}(\ket{1,\alpha}_\mathrm{T}\ket{0,\alpha'}_\mathrm{R}+
                                \ket{0,\alpha}_\mathrm{T}\ket{1,\alpha'}_2)\\
&=&D_{b_\mathrm{T}}(\alpha)D_{b_\mathrm{R}}(\alpha')
   \frac{1}{\sqrt{2}}(\Adagger_\mathrm{T}+\Adagger_\mathrm{R})
   \ket{0,0}_\mathrm{T}\ket{0,0}_\mathrm{R}\nonumber\\
&\rightarrow& D_{c_\mathrm{T}}(r\alpha)D_{d_\mathrm{T}}(t\alpha)
             D_{c_\mathrm{R}}(r\alpha')D_{d_\mathrm{R}}(t\alpha')\times\nonumber\\
&& (t\hat{c}^{\dag}_\mathrm{T}-r\hat{d}^{\dag}_\mathrm{T}+
 t\hat{c}^{\dag}_\mathrm{R}-r\hat{d}^{\dag}_\mathrm{T})
 \ket{0,0}_\mathrm{T}\ket{0,0}_\mathrm{R}\nonumber\\
&=&(t\hat{c}^{\dag}_\mathrm{T}-r\hat{d}^{\dag}_\mathrm{T}+
    t\hat{c}^{\dag}_\mathrm{R}-r\hat{d}^{\dag}_\mathrm{R})
    \ket{r\alpha,t\alpha}_\mathrm{T}\ket{r\alpha',t\alpha'}_\mathrm{R}\nonumber,
\end{eqnarray}
where $\ket{0,0}_\mathrm{T}\ket{0,0}_\mathrm{R}$ is the vacuum
state of the four modes. The probability for the two detectors to
click simultaneously is given by:
\begin{equation}
\label{eq:prob} \mathrm{P_{coincidence}}=1-(D_1+D_2-D_{12})
\end{equation}
where $D_1$ ($D_2$) is the probability that detector $1$ ($2$)
registers nothing irrespective of what happens in the other
detectors. The term  $D_{12}$ balances for the double count of the
event "nothing in both detectors". We use this measure instead of
photon number correlation because our detectors only register the
presence of photons and are unable to resolve the photon number.

To calculate these different probabilities we use the following
projection operators:
\begin{eqnarray*}
\hat{\mathrm{P}}_{1}&=&
|0\rangle\langle 0|\otimes\mathbf{1}\otimes\mathbf{1}\otimes\mathbf{1}\\
\hat{\mathrm{P}}_{2}&=&
\mathbf{1}\otimes\mathbf{1}\otimes |0\rangle\langle 0|\otimes\mathbf{1}\\
\hat{\mathrm{P}}_{12}&=&|0\rangle\langle
0|\otimes\mathbf{1}\otimes|0\rangle\langle 0|\otimes\mathbf{1},
\end{eqnarray*}
where the spaces refer, in order to the modes $c_\mathrm{T}$,
$d_\mathrm{T}$, $c_\mathrm{R}$, and $d_\mathrm{R}$.
Using the above definitions of $\hat{\mathrm{P}}_i$ and
$\ket{\Psi}$ one finds:
\begin{eqnarray*}
D_1&=&\bra{\Psi}\hat{\mathrm{P}}_{1}\ket{\Psi}=
    \frac{1}{2}e^{-|r\alpha|^2}(1+r^2+r^2t^2|\alpha|^2).\\
D_2&=&\bra{\Psi}\hat{\mathrm{P}}_{2}\ket{\Psi}=
    \frac{1}{2}e^{-|r\alpha|^2}(1+r^2+r^2t^2|\alpha|^2).\\
D_{12}&=&\bra{\Psi}\hat{\mathrm{P}}_{12}\ket{\Psi}=
e^{-2|r\alpha|^2}(r^2+2r^2t^2|\alpha|^2\cos^2\frac{\theta-\theta'}{2}).
\end{eqnarray*}
This yields the coincidence probability:
\begin{eqnarray}
&&\mathrm{P_{coincidence}}=
1-e^{-|r\alpha|^2}(1+r^2+r^2t^2|\alpha|^2)\nonumber\\
&&+e^{-2|r\alpha|^2}(r^2+2r^2t^2|\alpha|^2\cos^2\frac{\theta-\theta'}{2}).
\label{eq:coinc}
\end{eqnarray}
In addition to these coincidences, we also have the case with
\emph{zero} photons arriving in the single photon channel.
Coincidences are registered also in this case when two photons
from the LO are detected. The probability for these false
coincidence counts is easily calculated in the same way as above:
\begin{eqnarray}
\mathrm{P_{coincidence}^{\mathrm{false}}}&=&
(1-e^{-|r\alpha|^2})^2
 \label{eq:coincfalse}
\end{eqnarray}
This may also be verified easily through a different reasoning:
The probability of having more than zero LO photons in one arm is
given by $1-e^{-|r\alpha|^2}$. The probability of having photons
in both arms is this probability squared.

The total probability for the two detectors to register photons
simultaneously is given by
$$
\mathrm{P_{coincidence}^{\mathrm{tot}}}=
\eta\mathrm{P_{coincidence}}+
(1-\eta)\mathrm{P_{coincidence}^{\mathrm{false}}}
$$
where $\eta$ is the quantum efficiency of the triggered photon
source (For the setup: $\eta\sim 10^{-2}$). To minimize the
influence of the false coincidences we choose $r\ll t$ and
$|r\alpha|$ small to minimize the influence of
$\mathrm{P_{coincidence}^{\mathrm{false}}}$. This choice
introduces a trade-off between the implementation of the
projectors described by Eq.~(\ref{eq:projstate}) and the goal to
minimize the influence of false coincidences. Practically, the
lower bound of $|r\alpha|$ is determined by the detection rate
allowed by the laser stability.

\begin{figure}[htb]
 \infig{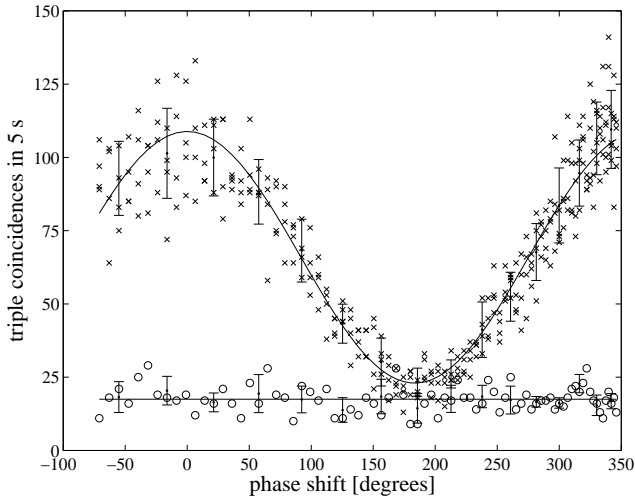}{8.5cm}
   \caption{Experimental data and curve fits. The oscillating
   curve shows the correlation between the two detectors as the
   phase shift is varied. The flat curve shows the correlation
   background due to multiple photons from the single photon
   source and the local oscillator. The visibility for the
   correlation curve is $91\pm 3\%$ and $66\pm 2\%$ with and
   without corrections for background correlations.
   These visibilities should be compared with the 71\% limit for
   violation of Bells inequalities.
   }
   \label{fig:data}
\end{figure}

In Fig.~\ref{fig:data} we plot the measured correlation obtained
in the setup illustrated in Fig.~\ref{fig:setup}. The birefringent
plate is rotated around the optic axis so that relative phase
shifts between -70 and 350 degrees are introduced in the two arms.
If either the single photon or the local oscillator is missing, we
measure a flat, phase independent correlation curve. Similarly,
the signal intensity in each detector is almost constant as the
phase shift is varied. Introducing a quarter-wave plate, with the
optical axis horizontal or vertical, in either
arm shifts the fringe pattern 90 degrees with negligible loss in
visibility.

This experiment is limited by multiple down-conversion events
since we chose to work at high UV-intensities (350 mW) to reduce
measurement times. With the local oscillator blocked, we detect
about one triple coincidence per second due to multiple
down-conversion pair production. Reduced intensity increases
visibility, see Pittman and Franson \cite{Pittman2003} for
details on this topic.

In our experiment we have raw correlation data with a visibility
$66\pm 2\%$. This curve consists of three contributions: i) The
two photons arriving to detectors $D_1$ and $D_2$ originate from
LO and single photon source (phase shift dependent). ii) The two
photons both come from the LO (constant under phase shift: ~2.5
$s^{-1}$). iii) The two photons both come from the single photon
source (constant under phase shift: ~1.0 $s^{-1}$). If the
coincidences corresponding to these two backgrounds are subtracted
from the signal, the visibility becomes $91\pm 3\%$.

It should be noted that it is possible to use different photon
states as the input state to the beam splitter (BS in
Fig.~\ref{fig:setup}). If the state remains separable after the
beam splitter, only classical correlations are expected. We
performed such tests using phase modulated coherent light instead
of single photons. Theoretically we would expect the correlation
to reach the maximally allowed 50 \% limit for this classical
correlation. Experimentally we create this coherent field by
splitting off light from the local oscillator to the polarization
control of the single photon channel (See Fig.~\ref{fig:setup}).
The phase modulation is provided by another delay line modifying
the propagation distance. Using this setup we measured a
correlation visibility of $48\pm 2\%$, just below the classical
limit. This visibility indicates a well aligned system and that we
don't observe Bell-type correlations in the local oscillator.

The representation of quantum information as a superposition
particle number states, $\alpha\ket{0}+\beta\ket{1}$ instead of
superimposing different modes offers certain advantages. Using
this representation makes it is feasible to perform quantum
computations with linear operations and feedback from measurements
\cite{Knill2001}. To perform interesting experiments on these
qubits, it is desirable to have access to precise quantum
mechanical observables in this space and specifically those with
eigenstates that are superpositions of number states
($\ket{0}+e^{i\theta}\ket{1}$). Here we have presented a robust
non-interferometric method for the implementation of such
observables. This experimental scheme may be scaled up to include
perform correlation measurements on multiphoton states. Using the
criteria for single particle non-locality set up by Hardy
\cite{Hardy1995reply}, we have performed an experiment that
supports the prediction of Tan \emph{et al.} of single particle
non-locality~\cite{Tan1991}.

We gratefully acknowledge useful discussions with Drs. Marie Ericsson,
Per Jonsson and Phil Marsden. This work was financially supported
by Swedish Research Council (VR), STINT and INTAS.
\bibliography{bell}
\end{document}